\documentclass[twocolumn,prb,showpacs,preprintnumbers,amsmath,amssymb]{revtex4}
\usepackage{graphicx}
\begin{document}
\title{Spin polarized transport in the weak link between f-wave superconductors}
\author{Gholamreza Rashedi$^{1}$, Yuri A. Kolesnichenko$^{2}$}
\address{$^1$ Institute for Advanced Studies in Basic Sciences, 45195-159,
Zanjan, Iran\\ $^2$ B.Verkin Institute for Low Temperature Physics
 Engineering of National Academy of Sciences of Ukraine, 47,
  Lenin ave , 61103, Kharkov, Ukraine}
\date{\today}
\begin{abstract}
The spin current in the Josephson junction as a weak-link
(interface) between misorientated triplet superconductors is
investigated theoretically for the models of the order parameter
in $UPt_{3}$. Green functions of the system are obtained from the
quasiclassiacl Eilenberger equations. The analytical results for
the charge and spin currents are illustrated by numerical
calculations for the certain misorientation angles of gap vector
of superconductors. As the main result of this paper, it is found
that, at some values of the phase difference, at which the charge
current is exactly zero, the spin current has its maximum value.
Furthermore, it is shown that the origin of spin current is the
misorientation between gap vectors of triplet superconductors.
\end{abstract}
\maketitle
\section{Introduction}
Triplet superconductivity has become one of the most interesting
topics of condensed matter physics \cite{Maeno,Mackenzie},
particularly in view of the recently discovered ferromagnetic
superconductivity \cite{Mineev,Samokhin}. The mechanism of
pairing, physics of interaction and gap structure in this type of
superconductors have been the subject of many experimental and
theoretical works \cite
{Ishida1,Takimoto,Asano,Karchev,Deguchi,Blount,Anderson}. The
Cooper pairing in the triplet superconductors has been
investigated, for example, using the thermal conductivity in
Refs.\cite{Izawa,Graf} and Knight shift experiments in Refs.
\cite{Tou,Ishida2}. Also, the Josephson effect in the point
contact between triplet superconductors has been studied in
Ref.\cite{Mahmoodi}. These weak-link structures have been used to
demonstrate the order parameter symmetry in Ref.\cite{Stefanakis}.
Eventually, the $f$-wave
symmetry of order parameter has been proposed for $UPt_{3}$ and $%
Sr_{2}RuO_{4}$ compounds. In addition, the spin polarized
transport through the systems consisting of superconductors,
normal metals, ferromagnetic layers and other structures as one of
the modern topics of mesoscopic physics, has attracted much
attention recently \cite {Imamura,Chtchelkatchev,Sun,Maekawa}. In
the present paper, the ballistic Josephson weak-link as the
interface between two bulk of $f$-wave superconductors with
different orientations of the crystallographic axes has been
investigated. It is shown that the current-phase dependencies are
totally different from the current-phase dependencies of the
junction between conventional ($s$-wave) superconductors
\cite{Kulik,Rashedi} and high $T_{c}$ ($d$-wave) superconductors
\cite{Coury}. It is found that for the certain values of the
misorientation, the spin-current in the both directions,
tangential and perpendicular to the interface, may exist and it
has totally unusual dependence on the external phase difference.
The effect of misorientation on the spin current is investigated.
It is observed that the misorientation between gap vectors is the
origin of the spin current. As the important result of this paper,
it is obtained that, at some of certain values of phase
difference, at which the charge current is zero, the spin current
has the finite value. Another result of the paper is the
capability of this proposed experiment for polarization of the
spin transport using the junction between $f$-wave
superconductors. Eventually, one of the states and geometries of
our system can be used as a switch which is able to divide the
spin and charge currents into two
parts: parallel and perpendicular to the interface.\\
The organization of the rest of this paper is as follows. In
Sec.$\text{II}$ we describe our configuration, which is
investigated. For a non-self-consistent model of the order
parameter, the quasiclassiacl Eilenberger
equations\cite{Eilenberger} are solved and suitable Green
functions are obtained analytically. In Sec. $\text{III}$ the
obtained formulas for the Green functions are used for calculation
the charge and spin current densities at the interface. An
analysis of numerical results will be done in Sec.$\text{IV}$. The
paper will be finished with some conclusions in Sec.$\text{V}$.
\section{Formalism and Basic Equations}
We consider a model of a flat interface $y=0$ between two
misorientated $f-$wave superconducting half-spaces (Fig.1) as a
ballistic Josephson junction. In the quasiclassical ballistic
approach, in order to calculate the charge and spin current, we
use ``transport-like'' equations\cite{Eilenberger}
for the energy integrated Green functions $\breve{g}\left( \mathbf{\hat{v}}_{F},%
\mathbf{r},\varepsilon _{m}\right) $
\begin{equation}
\mathbf{v}_{F}\nabla \breve{g}+\left[ \varepsilon _{m}\breve{\sigma}_{3}+i%
\breve{\Delta},\breve{g}\right] =0,  \label{Eilenberger}
\end{equation}
and the normalization condition
\begin{equation}
\breve{g}\breve{g}=\breve{1},  \label{Normalization}
\end{equation}
where $\varepsilon _{m}=\pi T(2m+1)$ are discrete Matsubara energies $%
m=0,1,2...$, $T$ is the temperature and $\mathbf{v}_{F}$ is the Fermi velocity and $\breve{\sigma}_{3}=%
\hat{\sigma}_{3}\otimes \hat{I}$ in which $\hat{\sigma}_{j}\left(
j=1,2,3\right) $ are Pauli matrices.
\begin{figure}[tbp]
\includegraphics[width=\columnwidth]{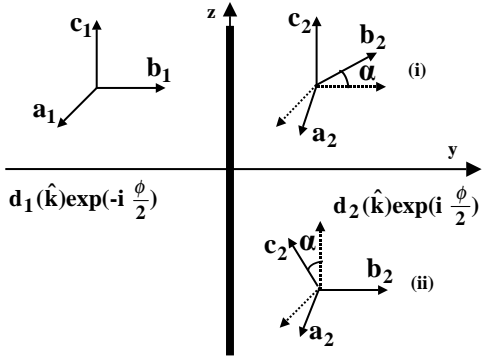}
\caption{Scheme of a flat interface between two superconducting
bulks, which are misorientated as much as $\protect\alpha $.}
\label{fig1}
\end{figure}
The Matsubara propagator $\breve{g}$ can be written in the
standard form:
\begin{equation}
\breve{g}=\left(
\begin{array}{cc}
g_{1}+\mathbf{g}_{1}\mathbf{\hat{\sigma}} & \left( g_{2}+\mathbf{g}_{2}\hat{%
\mathbf{\sigma }}\right) i\hat{\sigma}_{2} \\
i\hat{\sigma}_{2}\left( g_{3}+\mathbf{g}_{3}\hat{\mathbf{\sigma
}}\right)  &
g_{4}-\hat{\sigma}_{2}\mathbf{g}_{4}\hat{\mathbf{\sigma
}}\hat{\sigma}_{2}
\end{array}
\right) ,
\end{equation}
\label{Green's function} where, the matrix structure of the off-diagonal self energy $\breve{%
\Delta}$ in the Nambu space is
\begin{equation}
\breve{\Delta}=\left(
\begin{array}{cc}
0 & \mathbf{d}\hat{\mathbf{\sigma }}i\hat{\sigma}_{2} \\
i\hat{\sigma}_{2}\mathbf{{d^{\ast }}\hat{\sigma}} & 0
\end{array}
\right) .
\end{equation}
\label{order parameter} In this paper, the unitary states, for which $\mathbf{%
d\times d}^{\ast }=0,$ is investigated. Also, the unitary states
vectors $\mathbf{d}_{1,2}$ can be written as
\begin{equation}
\mathbf{d}_{n}=\mathbf{\Delta }_{n}\exp i\psi _{n},
\end{equation}
where $\mathbf{\Delta }_{1,2}$ are the real vectors in the left
and right sides of the junction.\ The gap (order parameter) vector
$\mathbf{d}$ has to be determined from the self-consistency
equation:
\begin{equation}
\mathbf{d}\left( \mathbf{\hat{v}}_{F},\mathbf{r}\right) =\pi
TN\left(
0\right) \sum_{m}\left\langle V\left( {\mathbf{\hat{v}}}_{F},{\mathbf{\hat{v}%
}}_{F}^{\prime }\right) \mathbf{g}_{2}\left(
{\mathbf{\hat{v}}}_{F}^{\prime },\mathbf{r},\varepsilon
_{m}\right) \right\rangle   \label{self-consistent}
\end{equation}
where $V\left(
{\mathbf{\hat{v}}}_{F},{\mathbf{\hat{v}}}_{F}^{\prime }\right) $,
is a potential of pairing interaction, $\left\langle
...\right\rangle $ stands for averaging over the directions of an
electron momentum on the Fermi surface
${\mathbf{\hat{v}}}_{F}^{\prime }$ and $N\left( 0\right) $ is the
electron density of states at the Fermi level of energy.
\begin{figure}[tbp]
\includegraphics[width=\columnwidth]{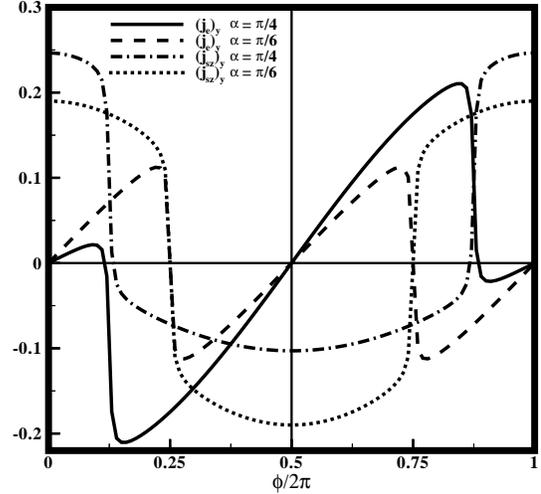}
\caption{Charge and spin current ($s_{z}$) versus the phase
difference $\protect\phi $ for the planar state (\ref{planar}),
geometry (i) and the different misorientations. Currents are given
in units of $j_{0,e}=\frac{\protect\pi }{2}eN(0)v_{F}\Delta
_{0}(0) $ and $j_{0,s}=\frac{\protect\pi }{4}\hbar N(0)v_{F}\Delta
_{0}(0)$ respectively.} \label{fig2}
\end{figure}
Solutions to Eqs. (\ref{Eilenberger}) and (\ref{self-consistent})
must satisfy the conditions for Green functions and vector
$\mathbf{d}$ in the bulks of the superconductors far from the
interface as follow:
\begin{eqnarray}
\breve{g}\left( \pm \infty \right)  &=&\frac{\varepsilon _{m}\breve{\sigma}%
_{3}+i\breve{\Delta}_{2,1}}{\sqrt{\varepsilon _{m}^{2}+\left| \mathbf{d}%
_{2,1}\right| ^{2}}};  \label{Bulk solution} \\
\mathbf{d}\left( \pm \infty \right)  &=&\mathbf{d}_{2,1}\left( \mathbf{\hat{v%
}}_{F}\right) \exp \left( \mp \frac{i\phi }{2}\right) ,
\label{Bulk order parameter}
\end{eqnarray}
where $\phi $ is the external phase difference between the order
parameters of the bulks. Eqs. (\ref{Eilenberger}) and
(\ref{self-consistent}) have to be supplemented by the continuity
conditions at the interface between superconductors. For all
quasiparticle trajectories, the Green functions satisfy the
boundary conditions both in the right and left bulks as well as at
the interface.\\
The system of equations (\ref{Eilenberger})and
(\ref{self-consistent}) can be solved only numerically. For
unconventional superconductors such solution
requires the information of the function $V\left( {\mathbf{\hat{v}}}_{F},{%
\mathbf{\hat{v}}}_{F}^{\prime }\right)$. This information, as that
of the nature of unconventional superconductivity in novel
compounds, in most cases is unknown. Usually, the spatial
variation of the order parameter and its dependence on
the momentum direction can be separated in the form of $\Delta ({\mathbf{%
\hat{v}}}_{F},y)=\Delta ({\mathbf{\hat{v}}}_{F})\Psi (y)$. It has
been shown that the absolute value of a self-consistent order
parameter and $\Psi (y)$ are suppressed near the interface and at
the distances of the order of the coherence length, while its
dependence on the direction in the momentum space ($\Delta
({\mathbf{\hat{v}}}_{F})$) remains unaltered \cite{Barash}.
\begin{figure}[tbp]
\includegraphics[width=\columnwidth]{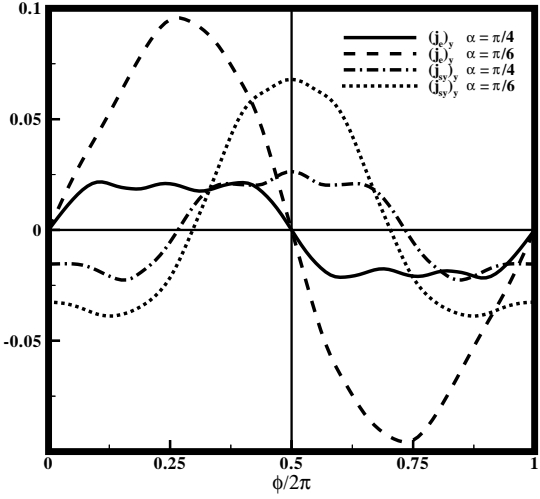}
\caption{Charge and spin current ($s_{y}$) versus the phase
difference $\protect\phi $ for
the axial state (\ref{axial}), geometry (ii) and the different misorientations ($y$%
-component).} \label{fig3}
\end{figure}
Consequently, this suppression doesn't influence the Josephson
effect drastically. This suppression of the order parameter keeps
the current-phase dependence unchanged but, it changes the
amplitude value of the current. For example, it has been verified
in Refs. \cite{Coury} for the junction between unconventional
$d$-wave, in Ref.\cite {Barash} for the case of ``$f$-wave''
superconductors and in Ref.\cite{Viljas} for pinholes in $^{3}He$
that, there is a good qualitative agreement between
self-consistent and non-self-consistent results. Also, it has been
observed that the results of the non-self-consistent investigation
of $D-N-D$ structure in the paper\cite{Faraii} are coincident with
the experimental results of the paper \cite{Freamat} and the
results of the non-self-consistent model in paper\cite{Yip} are
similar to the superfluid weak-link experiment\cite{Backhaus}.
Note that any self-consistent solution can be found for some
models of pairing potential and the certain values of parameters.
Consequently, despite the fact that this estimation cannot be
applied directly for a quantitative analyze of the real
experiment, only a qualitative comparison of calculated and
experimental current-phase relations is possible. In our
calculations, a simple model of the constant order parameter up to
the interface is considered and the pair breaking and the
scattering on the interface are ignored. We believe that under
these strong assumptions our results describe the real situation
qualitatively. In the framework of such model, the analytical
expressions for the charge and spin current can be obtained for an
arbitrary form of the order parameter.
\section{Analytical results.}
The solution of Eqs. (\ref{Eilenberger}) and
(\ref{self-consistent}) allows us to calculate the charge and spin
current densities. The expression for the charge current is:
\begin{equation}
\mathbf{j}_{e}\left( \mathbf{r}\right) =2i\pi eTN\left( 0\right)
\sum_{m}\left\langle \mathbf{v}_{F}g_{1}\left( \mathbf{\hat{v}}_{F},\mathbf{r%
},\varepsilon _{m}\right) \right\rangle  \label{charge-current}
\end{equation}
and for the spin current we have:
\begin{equation}
\mathbf{j}_{s_{i}}\left( \mathbf{r}\right) =2i\pi (\frac{\hbar
}{2})TN\left(
0\right) \sum_{m}\left\langle \mathbf{v}_{F}\left( \mathbf{{\hat{e}}}_{i}%
\mathbf{g}_{1}\left( \mathbf{\hat{v}}_{F},\mathbf{r},\varepsilon
_{m}\right) \right) \right\rangle  \label{spin-current}
\end{equation}
where, $\mathbf{{\hat{e}}}_{i}\mathbf{=}\left( \hat{\mathbf{x}},\hat{%
\mathbf{y}},\hat{\mathbf{z}}\right) $. We assume that the order
parameter
does not depend on coordinates and in each half-space it equals to its value (%
\ref{Bulk order parameter}) far form the interface in the left or
right bulks. For such a model, the current-phase dependence of a
Josephson junction can be calculated analytically. It enables us
to analyze the main features
of current-phase dependence for the different models of the order parameter of ``%
$f$-wave'' superconductivity. The Eilenberger equations
(\ref{Eilenberger}) for Green functions $\breve{g}$, which are
supplemented by the condition of continuity of solutions across
the interface, $y=0$, and the boundary conditions at the bulks,
are solved for a non-self-consistent model of the order parameter
analytically. Two diagonal terms of Green matrix which determine
the current densities at the interface, $y=0$, are following.
\begin{figure}[tbp]
\includegraphics[width=\columnwidth]{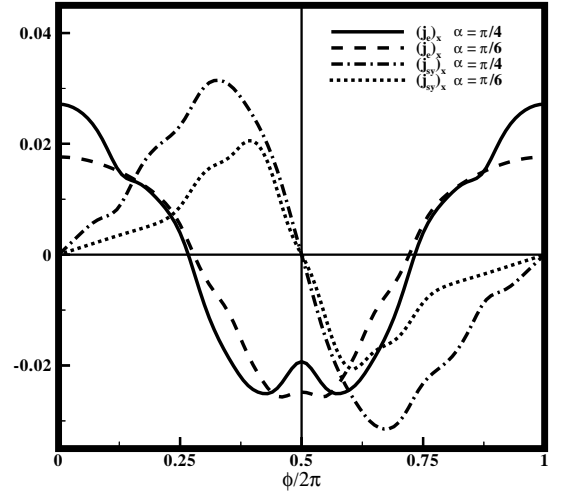}
\caption{Charge and spin current ($s_{y}$) versus the phase
difference $\protect\phi $ for
the axial state (\ref{axial}), geometry (ii) and the different misorientations ($x$%
-component)} \label{fig4}
\end{figure}
For the relative term to the charge current we obtain:
\begin{equation}
g_{1}\left( 0\right) =\frac{\varepsilon _{m}(\Omega _{1}+\Omega
_{2})\cos\beta+i\eta\sin\beta(\Omega _{1} \Omega _{2}+\varepsilon
_{m}^2)}{i\eta\sin\beta\varepsilon _{m}(\Omega _{1}+\Omega
_{2})+\cos\beta(\Omega _{1}\Omega _{2}+ \varepsilon
_{m}^2)+\mathbf{\Delta}_{1}\mathbf{\Delta}_{2}}
\label{charge-term}
\end{equation}
and for the case of spin current we have:
$$\mathbf{g_{1}}\left( 0\right)
=\frac{\eta\mathbf{\Delta}_{1}\times\mathbf{\Delta}_{2}}{(A+2B)\left|
\mathbf{d}_{1}\right| ^{2}\left| \mathbf{d}_{2}\right|
^{2}}[(B-1)^{2}\exp(i\beta)(\eta \Omega _{1}+\varepsilon
_{m})\times$$
\begin{equation}
(\eta \Omega _{2}+\varepsilon _{m})-(B+1)^{2}\exp(-i\beta)(\eta
\Omega _{2}-\varepsilon _{m})(\eta \Omega _{1}-\varepsilon
_{m})]\label{spin-term}
\end{equation}
where $\eta =sgn\left( v_{y}\right) $, $\Omega
_{n}=\sqrt{\varepsilon _{m}^{2}+\left| \mathbf{d}_{n}\right|
^{2}}$, $\beta=\psi_1-\psi_2+\phi$,
\begin{equation}
B=\frac{\eta\varepsilon _{m}(\Omega _{1}+\Omega
_{2})\cos\beta+i\sin\beta(\Omega _{1} \Omega _{2}+\varepsilon
_{m}^2)}{i\eta\sin\beta\varepsilon _{m}(\Omega _{1}+\Omega
_{2})+\cos\beta(\Omega _{1}\Omega _{2}+ \varepsilon
_{m}^2)+\mathbf{\Delta}_{1}\mathbf{\Delta}_{2}}
\label{mathematics-term}
\end{equation}
and
\begin{equation}
A=\frac{\mathbf{\Delta}_1\mathbf{\Delta}_2(B-1)\exp(i\beta)}{(\eta
\Omega _{1}-\varepsilon _{m})(\eta \Omega _{2}-\varepsilon
_{m})}+\frac{\mathbf{\Delta}_1\mathbf{\Delta}_2(B+1)\exp(-i\beta)}{(\eta
\Omega _{1}+\varepsilon _{m})(\eta \Omega _{2}+\varepsilon _{m})}
\end{equation}
Also, $n=1,2$ label the left and right half-spaces respectively.
\begin{figure}[tbp]
\includegraphics[width=\columnwidth]{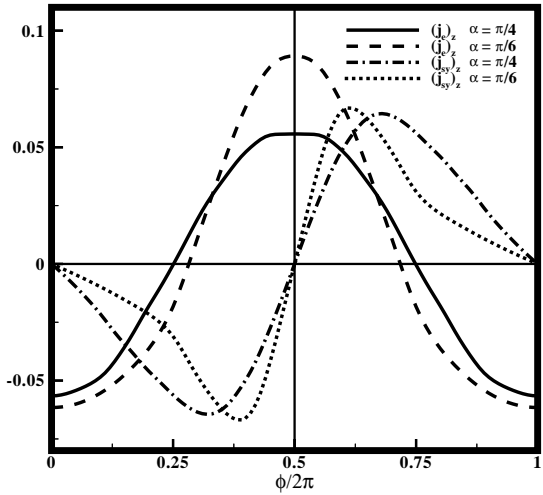}
\caption{Charge and spin current ($s_{y}$) versus the phase
difference $\protect\phi $ for
the axial state (\ref{axial}), geometry (ii) and the different misorientations ($z$%
-component)} \label{fig5}
\end{figure}
We consider a rotation $\breve{R}$ only in the right
superconductor (see, Fig.1), (i.e., $\mathbf{d}_{2}(\hat{\mathbf{k}}) =\breve{R}\mathbf{d}%
_{1}( \breve{R}^{-1}\hat{\mathbf{k}}),$ $\hat{\mathbf{k}}$ is the
unit vector in the momentum space). The crystallographic $c$-axis
in the left half-space is selected parallel to the partition
between the superconductors (along $z$-axis in Fig.1). To
illustrate the results obtained by computing the formula
(\ref{charge-term},\ref{spin-term}), we plot the
current-phase diagrams for the different models of the ``$f$-wave'' pairing symmetry (%
\ref{axial},\ref{planar}) and for two different geometries. These
two geometries are corresponding to the different orientations of
the crystals in the right and left sides of the interface (see,
Fig.1):\newline (i) The basal $ab$-plane in the right side has
been rotated around the $c$-axis by $\alpha $; $\hat{\mathbf{c}}_{1}\Vert \hat{\mathbf{c}}_{2}$.%
\newline
(ii) The $c$-axis in the right side has been rotated around the axis perpendicular to the interface ($y$%
-axis in Fig.1) by $\alpha $; $\hat{\mathbf{b}}_{1}\Vert
\hat{\mathbf{b}}_{2}$.\newline
Further calculations require a certain model of the gap vector (vector of order parameter) $%
\mathbf{d}$.
\section{Analysis of numerical results}
In the present paper, two most probable forms of the $f$-wave
order parameter vector in $UPt_{3}$ are considered. The first
model which is successful to explain the properties of the
$B$-phase of $UPt_{3}$ is the axial state. This sate describes the
strong spin-orbital coupling with vector \textbf{d} directed along
the \textbf{c} axis of the lattice and it is:
\begin{equation}
\mathbf{d}(T,\mathbf{v}_{F})=\Delta
_{0}(T)\hat{\mathbf{z}}k_{z}\left( k_{x}+ik_{y}\right) ^{2}.
\label{axial}
\end{equation}
The coordinate axes
$\hat{\mathbf{x}},\hat{\mathbf{y}},\hat{\mathbf{z}}$
here and below are chosen along the crystallographic axes $\hat{\mathbf{a}},%
\hat{\mathbf{b}},\hat{\mathbf{c}}$ in the left side of
Fig.\ref{fig1}. The function $ \Delta _{0}=$ $\Delta _{0}\left(
T\right) $ in Eq.(\ref{axial})
and below describes the dependence of the order parameter $\mathbf{d}$ on the temperature $%
T$ (our numerical calculations have been done at the temperatures
close to the $T=0$). The second model of the order parameter which
describes the weak spin-orbital coupling in $UPt_{3}$ states, is
the unitary planar state. The planar model of gap vector is:
\begin{equation}
\mathbf{d}(T,\mathbf{v}_{F})=\Delta
_{0}(T)k_{z}(\hat{\mathbf{x}}\left( k_{x}^{2}-k_{y}^{2}\right)
+\hat{\mathbf{y}}2k_{x}k_{y}). \label{planar}
\end{equation}
Using these two models of order parameters
(\ref{axial},\ref{planar}) and solutions to the Eilenberger
equations (\ref{charge-term},\ref{spin-term}), we have calculated
the spin current and charge current densities at the interface
numerically.
\begin{figure}[tbp]
\includegraphics[width=\columnwidth]{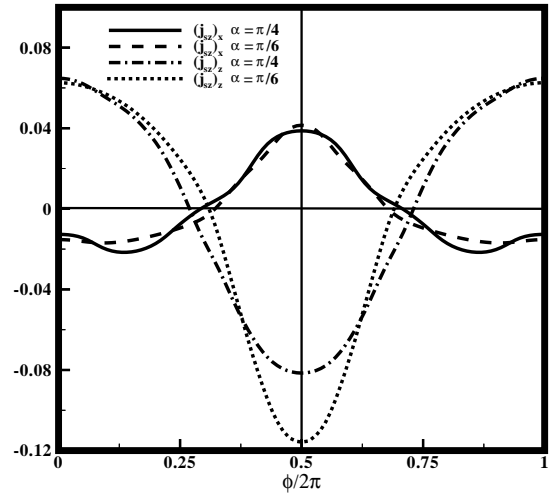}
\caption{Tangential spin current ($s_{z}$) versus the phase
difference $\protect\phi $ for the planar state (\ref{planar}),
geometry (ii) and the different misorientations. The perpendicular
component ($y$-direction) of the spin current is absent.}
\label{fig6}
\end{figure}
These numerical results are listed below:\newline 1) The spin
current can be present, only when misorientation between gap
vectors exists. Because in our Green function (\ref{spin-term}),
the spin current is proportional to the ``cross product'' between
the left and right gap vectors. For instance, the spin current for
the case of the axial state (\ref{axial}) and
geometry (i) is zero, because both of the gap vectors are in the same direction (%
$\mathbf{{\hat{z}}}$). (Geometry (i) is a rotation as much as
$\alpha $, around the $z$ axis).\newline 2) In Fig.\ref{fig2} it
is shown that for the planar state and geometry (i), it is
possible to observe the current of $s_{z}$ in the direction
perpendicular to the interface, but in Figs.\ref{fig3},\ref{fig4}
and \ref{fig5}, it is demonstrated that, for the axial state and
geometry (ii), only the current of $s_{y}$ can be observed.
\begin{figure}[tbp]
\includegraphics[width=\columnwidth]{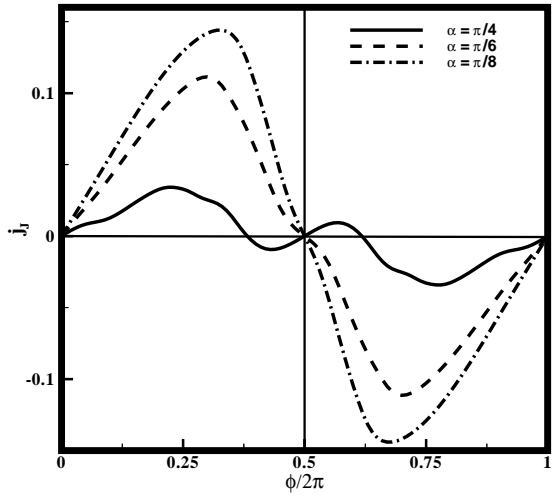}
\caption{The perpendicular component of the charge current (Josephson) versus the phase difference $%
\protect\phi $ for the planar state (\ref{planar}), geometry (ii)
and the different misorientations. The tangential components ($x$
and $z$-directions) are absent.} \label{fig7}
\end{figure}
Consequently, this kind of junction can be applied as a polarizer
or filter for the spin currents.\\ However, for the planar state
and geometry (ii), all terms of the spin current ($s_{x}$, $s_{y}$
and $s_{z}$) can be observed (see Eq.\ref{spin-term}).\newline 3)
In Figs.\ref{fig3}-\ref{fig9} (planar states), it is shown that
the value of the phase differences in which the currents are in
the maxima, minima and zero values, are not very sensitive to the
misorientation angle $\alpha $, while the amplitude of maxima and
minima, are strongly dependent on the value of misorientation
$\alpha $.\\
4) In the Figs.\ref{fig2},\ref{fig3} and Fig.\ref{fig6},
\ref{fig7}, while the charge currents are the odd functions of
$\phi $ with respect to the line of $\phi =\pi $, the spin
currents are even functions of the phase difference; $\mathbf{j}_{e}(\phi =\pi +\delta \phi )=-\mathbf{j}%
_{e}(\phi =\pi -\delta \phi )$ and for the spin current $\mathbf{j}%
_{s_{i}}(\phi =\pi +\delta \phi )=\mathbf{j}_{s_{i}}(\phi =\pi
-\delta \phi ) $. On the contrary, in the Figs.\ref{fig4} and
\ref{fig5}, the
charge and spin currents are even and odd functions of $\phi $ with respect to the line of $%
\phi =\pi $, respectively; $\mathbf{j}_{e}(\phi =\pi +\delta \phi )=\mathbf{j%
}_{e}(\phi =\pi -\delta \phi )$ and $\mathbf{j}_{s_{i}}(\phi =\pi
+\delta \phi )=-\mathbf{j}_{s_{i}}(\phi =\pi -\delta \phi
)$.\newline 5) In Fig.\ref{fig2}, the perpendicular component of
the spin and charge current in terms of the external phase
difference $\phi $ for the case of the planar state
(\ref{planar}), geometry (i) and for two different misorientations
are plotted. The solid line is the charge current-phase dependence
\cite {Mahmoodi}. Also, at the $\phi =0$, $\phi =\pi $ and $\phi
=2\pi $, the charge current (Josephson current) is zero while the
spin current has the
finite value.\\
6) The perpendicular component of the charge (Josephson current)
and spin current for the case of the axial state (\ref{axial}) and
geometry (ii) are plotted in Fig.\ref{fig3}, and the tangential
components of them, are plotted in Figs.\ref{fig4},\ref{fig5}. The
charge current-phase diagrams have been obtained before in paper
\cite{Mahmoodi} and they are totally different from the case of
conventional superconductors in the paper\cite{Kulik}.
\begin{figure}
\includegraphics[width=\columnwidth]{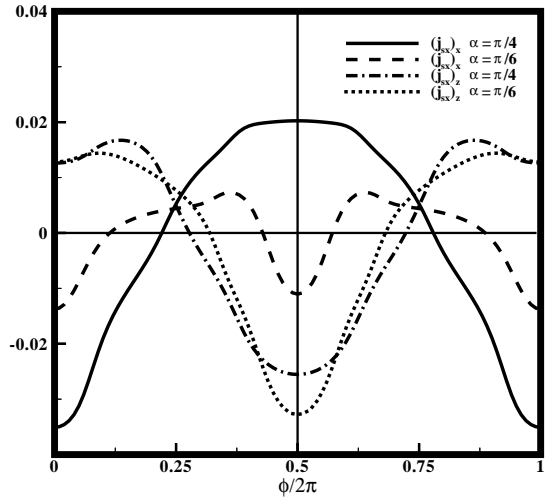}
\caption{Tangential spin currents ($s_{x}$) versus the phase
difference $\protect\phi $ for the planar state (\ref{planar}),
geometry (ii) and the different misorientations. The perpendicular
component ($y$-direction) of the spin current is absent.}
\label{fig8}
\end{figure}
At the phase values of $\phi =0$, $\phi =\pi $ and $\phi =2\pi $,
in which the charge current is exactly zero, the spin current has
the finite values and may select its maximum value. In
Figs.\ref{fig2},\ref{fig3} and specially Fig.\ref{fig9}, for a
small value of misorientation we have a very
long but narrow peak in the spin current phase diagram, close to the $\phi =\pi$.\\
7) Both the planar state with geometry (i) and the axial state
with geometry (ii) can be applied as the filter for polarization
of the spin transport (see Figs\ref{fig2}-\ref{fig5}), the former
transports only the $s_z$ but the latter case flows the $s_y$ (see
.\ref{spin-term}). In addition, the planar states with geometry
(ii) can be used as a switch for the spin and charge current into
two directions: parallel and perpendicular to the interface. In
this case, the spin and charge currents select only one of the
directions parallel or perpendicular to the interface. Namely, it
is impossible to observe the tangential and perpendicular
components of the currents at the same time for planar state with
geometry (ii)(Figs.\ref{fig6}-\ref{fig9}).
\section{Conclusions}
We have theoretically studied the spin current in the ballistic
Josephson junction in the model of an ideal transparent interface
between two misorientated $f$-wave superconductors which are
subject to a phase difference $\phi $. Our analysis has shown that
the misorientation and different models of the gap vectors
influence the spin current. This has been shown for the charge
current in the paper\cite{Mahmoodi}. The misorientation changes
strongly the critical values of both the spin current and charge
current. It has been obtained that the spin current is the result
of the misorientation between the gap vectors. Furthermore, it is
observed that the different models of the gap vectors and
geometries can be applied to the polarization of the spin
transport.
\begin{figure}[tbp]
\includegraphics[width=\columnwidth]{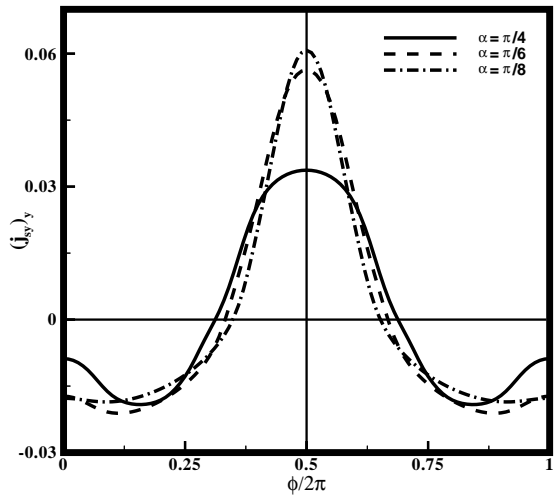}
\caption{Perpendicular component of the spin current ($s_{y}$) versus the phase difference $\protect%
\phi $ for the planar state (\ref{planar}), geometry (ii) and the
different misorientations. The tangential components ($x$ and $z$
directions) are absent.} \label{fig9}
\end{figure}
Another result of these calculations is the state in which the
currents select one of the two possible directions ( perpendicular
and parallel to the interface) to flow. This property can be used
as a switch to control the direction of the charge and spin
current. Finally, as an interesting and new result, it is observed
that at some certain values of the phase difference $\phi $, the
charge-current vanishes while the spin-current flows, although the
carriers of both spin and charge are the same (electrons). The
spatial variation of the phase of the order parameter plays a role
as the origin of the charge current and, similarly, due to the
broken $G^{spin-orbit}$ symmetry, a spatial difference of the gap
vectors in two half-spaces causes spin currents. This is because
there is a position-dependent phase difference between ``spin up''
and ``spin down'' Cooper pairs and, although the total charge
current vanishes, there can be a net transfer of the spin.
Therefore, in our system, there is a discontinuous jump between
the gap vectors and, consequently the spin currents should
generally be present. For instance, if up-spin states \ and
down-spin states have a velocity in the opposite direction, the
charge currents cancel each other whereas the spin current is
being transported. Mathematically
speaking, $\mathbf{{j_{charge}}={j_{\uparrow }}+{j_{\downarrow }},  {j_{spin}}=%
{j_{\uparrow }}-{j_{\downarrow }}}$, so it is possible to find the
state in which one of these current terms is zero and the other
term has the finite value\cite{Maekawa}. In addition the spin
imbalance which is the result of the different density of states
for ``spin-up'' and ``spin-down'' can be the other reason of spin
current\cite{Sun}. In conclusion, the spin current in the absence
of the charge current can be observed.

\end{document}